\documentclass[conference]{IEEEtran}
\IEEEoverridecommandlockouts
\usepackage{hyperref}
\hypersetup{
 pdfauthor={Md Saiful Islam, Rahul Bhadani},   % Author of the document
 pdftitle={Resilient Composite Control for Stability Enhancement in EV Integrated DC Microgrids},   % Title of the document
 pdfsubject={Power Systems, Control Systems, Power Electronics}, % Subject of the document
 pdfkeywords={DC microgrid, constant power load, virtual
capacitor, composite controller},
 pdfcreator={Md Saiful Islam}, % Creator
 pdfproducer={Md Saiful Islam}, % PDF producer
 colorlinks=true,
 filecolor=green, 
 citecolor = blue,       
 urlcolor=red, 
 }
 
\usepackage{cite}
\usepackage{amsmath,amssymb,amsfonts}
\usepackage{algorithmic}
\usepackage{graphicx}
\usepackage{textcomp}
\usepackage{xcolor}
\def\BibTeX{{\rm B\kern-.05em{\sc i\kern-.025em b}\kern-.08em
    T\kern-.1667em\lower.7ex\hbox{E}\kern-.125emX}}
\begin{document}

\title{Resilient Composite Control for Stability Enhancement in EV Integrated DC Microgrids\\
{\footnotesize \textsuperscript{}}
%\thanks{Identify applicable funding agency here. If none, delete this.}
}

\author{\IEEEauthorblockN{ Md Saiful Islam}
\IEEEauthorblockA{\textit{AI, Autonomy, Resilience, Control (AARC) Lab}\\\textit{Electrical \& Computer Engineering} \\
\textit{The University of Alabama in Huntsville}\\
Huntsville, Alabama, USA \\
mi1499@uah.edu}
\and

\IEEEauthorblockN{Rahul Bhadani}
\IEEEauthorblockA{\textit{AI, Autonomy, Resilience, Control (AARC) Lab}\\\textit{Electrical \& Computer Engineering} \\
\textit{The University of Alabama in Huntsville}\\
Huntsville, Alabama, USA \\
rahul.bhadani@uah.edu}
\and
}

\maketitle

\begin{abstract}

When electric vehicles (EVs) are integrated into the standalone DC microgrids (DCMGs), stability issues arise due to their constant power load (CPL) behavior, which provides negative incremental impedance (NII). In addition, the microgrids suffer from an inherent low-inertia problem. Therefore, this study presents a composite controller incorporating a global integral terminal sliding mode controller with a backstepping controller. A virtual capacitor is employed to mitigate the lower inertia issue and strengthen the DC-bus response. An improved fractional power–based reaching law decreases chattering as well as accelerates convergence. Exact feedback linearization converts the nonlinear boost converter model into Brunovsky’s canonical form, thereby mitigating NII effects and non-minimum phase issues.  The entire system stability is verified using Lyapunov control theory. Finally, simulation outcomes confirm superior performance, with 34.4–53.3\% reduction in overshoot, 52.9–74.9\% in undershoot, and 12–47.4\% in settling time compared to the existing controller.

\end{abstract}

\begin{IEEEkeywords}
DC microgrid, constant power load, virtual capacitor, composite controller
\end{IEEEkeywords}

\section{Introduction}

%%%%%%%%%%%%%%%%%%%%%%%%%%%%%%%

In recent years, DC microgrids (DCMGs) have been recognized as an efficient substitute for traditional AC systems, since they directly supply electric power from renewable sources to DC loads except any intermediate AC-to-DC or DC-to-AC conversion. Such an arrangement decreases power losses and simplifies control effort by avoiding synchronization, reactive power, and harmonic issues~\cite{senapati2024advancing}. Among several applications, DCMGs are rapidly employed for electric vehicle (EV) charging. However, integrating EVs into the DC-bus introduces critical stability challenges~\cite{rahman2024stability}. Owing to the lack of synchronous machines, DCMGs inherently suffer from low inertia degrading system dynamics. Moreover, EVs act as constant power loads (CPLs), providing negative incremental impedance (NII)~\cite{dubey2013understanding, tian2021electric}. This behavior induces oscillations, destabilization of the DC-bus voltage, and leads to system instability, during transient conditions with power mismatch and external disturbances~\cite{palash2025design}. To alleviate these challenges, nonlinear controllers combined with the virtual capacitors theory might be a good candidate to enhance voltage regulation and ensure robust, stable operation of EV-integrated DCMGs.

\subsection{Background}
\vspace{-5pt}

Various controllers have been used to enhance the stability of a microgrid with an EV charging station. The model predictive controller (MPC) has the ability to handle multivariable systems such as EV-connected DCMG with constraints~\cite{sadiq2025decentralized}. However, the efficacy of MPC degrades under extreme operating conditions as it performs suboptimally under large disturbances resulting from CPLs. Furthermore, the intensive computation of MPC restricts practical deployment. Despite these, the non-minimum phase (NMP) behavior of the DC–DC boost converter (DBC) still causes microgrid destabilization. To address the NMP issue, a nonlinear feedback linearizing controller (NFBLC) has been developed while decreasing parameter uncertainty and noise sensitivity~\cite{ muheki2025energy}. Despite the theoretical merits, the NFBLC remains complex to deploy in practical scenarios due to the reliance on accurate parameter estimation, which is often infeasible in microgrids. To overcome the limitations of MPC and NFBLC, backstepping control (BSC) has been used in~\cite{sarrafan2022novel} to suppress the voltage fluctuation and chattering phenomena. However, the dependency on precise system models restricts robustness. 

Sliding mode control (SMC) scheme has been widely explored to enhance microgrid resilience and provide a high stability margin~\cite{mohammed2022sliding}. However, conventional SMC suffers from chattering and struggles to maintain global stability when CPLs are connected to the microgrid. Hence, to reduce chattering while improving robustness, several advanced SMC variants have been introduced in~\cite{hassannia2023robust, DISMC2025, ullah2024super, napole2021global}. In~\cite{hassannia2023robust}, robust SMC improves transient performance but suffers from chattering. To mitigate this issue, a double integral SMC~\cite{DISMC2025} has been designed to enhance steady-state robustness, but still falls short in addressing model uncertainties and external disturbances. The super twisting SMC~\cite{ullah2024super} features a continuous control surface that remarkably decreases chattering and reinforces robustness. But it exhibits slower convergence under fast transients. To guarantee global finite-time convergence, global integral terminal SMC~\cite{napole2021global} offers resilience and lower steady-state error. Despite this, moderate chattering remains an issue. 

By incorporating the merits of different controllers, composite controllers have been constructed to deal with DCMGs~\cite{islam2024RBITSMC}. In~\cite{palash2024designing}, a backstepping nonsingular fast terminal sliding mode controller (BNFTSMC) is proposed to stabilize the DC-bus voltage under CPL conditions. In~\cite{palash2025design}, an updated BNFTSMC is constructed incorporating an integral term into the controller design to reduce the steady-state error. However, they do not incorporate any virtual concept to emulate virtual inertia. To alleviate this issue, virtual capacitor-based composite controllers are designed for reinforcing the stability of DCMG~\cite{islam2025virtual} and hybrid AC-DC microgrid~\cite{islam2025enhanced}. However, CPLs are not employed there to test the robustness of the controllers. Therefore, it remains to investigate composite controllers with virtual capacitors for intensifying the stability of EV-integrated DCMG systems.

\subsection{Key Contributions}
\vspace{-5pt}
A composite nonlinear control strategy is proposed for EV-integrated DCMGs, combining system transformation, virtual energy emulation, and composite control to elevate DC-bus stability and dynamic efficacy. The main contributions are:

\begin{enumerate}
    \item Converting the nonlinear EV-integrated DCMG model to Brunovsky’s canonical form to address NMP and using the accumulated energy function to enhance damping.
    \item Incorporating a virtual capacitor to decrease voltage oscillations and compensate for low inertia.
    \item Developing a global integral terminal sliding mode backstepping controller (GITSMBC) with an improved fractional power reaching law to suppress chattering as well as facilitate fast and smooth response.
    \item Verifying system stability through Lyapunov control function and demonstrating enhanced voltage regulation, robustness, and smoother control through simulations.
\end{enumerate}

%%%%%%%%%%%%%%%%%%%%%%%%%%%%%
        \begin{figure}[!b]
        \centering \includegraphics[width=0.47\textwidth]{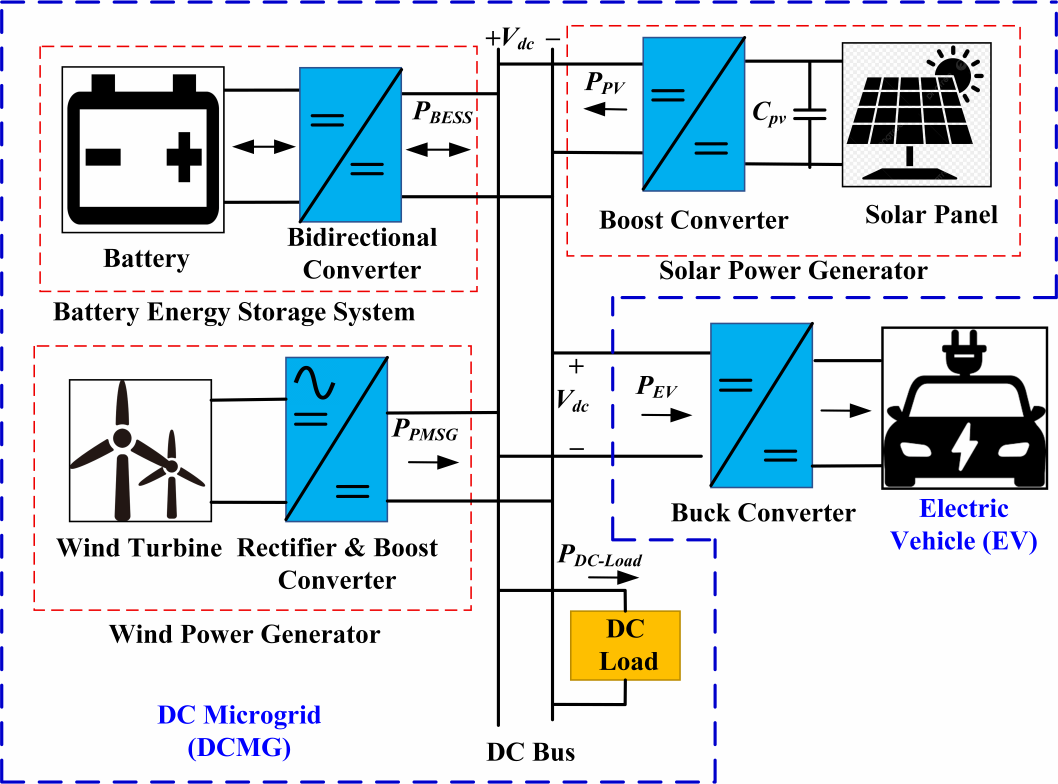}
        \caption{System schematic of the DC microgrid with electric vehicle integration.}
        \label{Fig1}
        \end{figure}
 %%%%%%%%%%%%%%%%%%%%%%%%%%%%%%%%%%%%

\section{Mathematical Modeling and Problem Formulation}

Figure~\ref{Fig1} illustrates the schematic of a DC microgrid (DCMG) consisting of solar PV, battery energy storage, wind turbine, DC–DC converters, and DC loads. An EV is connected to the DC-bus via a DC–DC converter, where it behaves as a CPL, while other DC loads act as constant voltage loads (CVLs). The CPL characteristics of EV introduces negative incremental impedance (NII), which threatens the stability margin of the DC-bus. A stable DC-bus voltage is essential for the reliable operation of the DCMG and fast charging of EVs. The operation of a CPL is governed by the constant power constraint as follows~\cite{palash2024designing}:
\begin{eqnarray}
\begin{aligned}\label{eqn1}
P_{EV} = P_{CPL} = V_{dc}(t) ~i_{CPL}(t) = \text{Constant}
\end{aligned}
\end{eqnarray}
Here, $P_{EV}$ or $P_{CPL}$ represents the EV's constant power demand, $V_{dc}(t)$ denotes the DC-bus voltage, and $i_{CPL}(t)$ is the current drawn by the CPL/EV. From this relationship, the CPL current can be derived as~\cite{palash2024designing}:
\begin{eqnarray}
\begin{aligned}\label{eqn2}
i_{EV} (t) = i_{CPL}(t) = \frac{P_{EV}}{V_{dc}(t)}
\end{aligned}
\end{eqnarray}
This nonlinear relation between $i_{CPL}(t)$ and $V_{dc}(t)$ results in an NII phenomenon, mathematically expressed as~\cite{palash2024designing}:
\begin{eqnarray}
\begin{aligned}\label{eqn3}
Z_{EV} = Z_{CPL} = \frac{dV_{dc}}{di_{CPL}} = -\frac{V_{dc}^2(t)}{P_{EV}}
\end{aligned}
\end{eqnarray}
Equation~(\ref{eqn3}) shows the NII of the CPL affects system dynamics and DC-bus stability, requiring the proposed controller to assure global stability. So, a nonlinear dynamical model of the DBC with CPLs is developed, as shown in Figure~\ref{Fig2}. A conventional way to reduce the CPL-induced instability is to increase the DC-bus capacitance ($C_{dc}$), since:
\begin{equation}\label{eqn16}
E = \cfrac{C_{dc} V_{dc}^2}{2} ,
P = \cfrac{dE}{dt} = C_{dc} V_{dc} \cfrac{dV_{dc}}{dt},
\cfrac{dV_{dc}}{dt} = \cfrac{P}{C_{dc} V_{dc}}
\end{equation}
Equation~\eqref{eqn16} shows that larger $C_{dc}$ slows voltage variations and enhances stability. But physical capacitors suffer from constraints of size, cost, and weight. To resolve this drawback, a virtual capacitor is introduced to emulate additional capacitance through control, acting in parallel with $C_{dc}$ as shown in Figure~\ref{Fig2}. The effective capacitance becomes:
\begin{equation}
C_{\text{eff}} = C_{dc} + C_{\text{vir}}
\end{equation}
in which $C_{dc}$ stands for virtual capacitor. The virtual capacitor supplies transient compensation power as:
\begin{equation}
\Delta P_{\text{vir}} = V_{dc} C_{\text{vir}} \cfrac{dV_{dc}}{dt}
\end{equation}
which improves damping and accelerates DC-bus voltage recovery. Incorporating the virtual capacitor concept, the system dynamics is written as follows:
\vspace{-10pt}
%%%%%%%%%%%%%%%%%%%%%%%%%%%%%
        \begin{figure}[!t]
        \centering \includegraphics[width=0.47\textwidth]{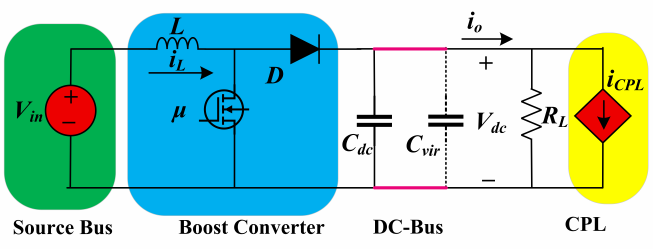}
        \caption{Simplified circuit diagram of the DC microgrid with electric vehicle integration.}
        \label{Fig2}
        \vspace{-18pt}
        \end{figure}
 %%%%%%%%%%%%%%%%%%%%%%%%%%%%%%%%%%%%

\begin{eqnarray}
\begin{aligned}\label{Eqn_7}
\frac{di_{L}}{dt} &= \frac{V_{\text{in}} - V_{dc}}{L} + \frac{V_{dc}}{L} \mu\\
\frac{dV_{dc}}{dt} &= \frac{1}{C_{\text{eff}}}\bigg(i_{L} - \frac{V_{dc}}{R_L} + \frac{\Delta P_{\text{vir}}}{V_{dc}} - \frac{P_{\text{CPL}}}{V_{dc}}\bigg) - \frac{i_{L}}{C_{\text{eff}}}\mu
\end{aligned}
\end{eqnarray}
Here, $V_{in}$ is the input voltage, $\mu$ denotes the switching control signal of the DBC, $i_{L}$ is the inductor current, $L$ is the DBC's inductance, $C_{dc}$ is the capacitance of the DC-bus, and $R_L$ represents the resistive load.
\\
\textbf{Remark 1.}~\textit{Virtual capacitor improves voltage stability by increasing $C_{\text{eff}}$. Yet, CPLs still introduce a destabilizing NII, requiring a composite control strategy. Since the system also exhibits NMP behavior, a canonical transformation is applied to decouple dynamics and enable effective controller design.}

%%%%%%%%%%%%%%%%%%%%%%%%

\section{Proposed Controller Design}
The main objective is to design a composite controller for a DBC under CPL/EV. However, the system dynamics, as presented in Equation~\eqref{Eqn_7}, is not useful to design a controller. Hence, the model should be modified to align with the requirements of the proposed composite controller.

\subsection{Conversion to Canonical Form}

The dynamical model of EV-integrated DCMG is converted into canonical form to overcome the NMP issue firstly. By applying an exact feedback linearization in Equation~\eqref{Eqn_7}, the system is then transformed into a completely linearized and decoupled form. This form is consistent with the Brunovsky canonical structure and is expressed as follows~\cite{palash2024designing}:
        \begin{eqnarray}
        \begin{aligned}\label{eqn32}
        \dot{\zeta}_1 & = \zeta_2\\
        \dot{\zeta}_2 & = a(x) + b(x) \mu
        \end{aligned}
        \end{eqnarray}
where $a(x) = \cfrac{V_{\text{in}} (V_{\text{in}} - x_2)}{L} - \cfrac{2x_2}{R_L C_{\text{eff}}} \left( x_1 - \cfrac{x_2}{R_L} + \cfrac{\Delta P_{C-\text{vir}}}{x_2} - \cfrac{P_{\text{CPL}}}{x_2} \right)$ and $b(x) = \cfrac{V_{\text{in}} x_2}{L_{}} + \cfrac{2 x_1 x_2}{R_L C_{\text{eff}}}$. Now, Equation~\eqref{eqn32} can be stated as follows incorporating the external disturbances:
        \begin{eqnarray}
        \begin{aligned}\label{eqn34}
        \dot{\zeta}_1 & = \zeta_2 + ED_1\\
            \dot{\zeta}_2 & = a(x) + b(x) \mu + ED_2
        \end{aligned}
        \end{eqnarray}
Here, $ED_1$ and $ED_2$ are called the external disturbances.
%%%%%%%%%%%%%%%%%%%%%%%%%%%

\subsection{Design Framework for the Composite Controller}

The following structured steps are employed to design the control law and fulfill the specified control objectives.

\begin{itemize}
    \item \textbf{Step 1}
\end{itemize}

The primary tracking error is
\begin{equation} \label{eqn41}
    e_1 = \zeta_1 - \zeta_{1(\text{ref})}
\end{equation}
The reference energy function $\zeta_{1(\text{ref})}$ is given by:
\begin{equation} \label{eqn411}
    \zeta_{1(\text{ref})} = \frac{1}{2} L i_{L(\text{ref})}^2 + \frac{1}{2} C_{\text{eff}} V_{\text{dc(ref)}}^2
\end{equation}
Here, $i_{L(\text{ref})} = \cfrac{P_{\text{CPL}}}{V_{\text{in}}}$ denotes the reference inductor current of the converter, and $V_{{dc(ref)}}$ represents the reference DC-bus voltage. Differentiating $e_1$ with respect to time:
\begin{equation} \label{eqn42}
    \dot{e}_1 = \zeta_2 + ED_1 - \dot{\zeta}_{1(\text{ref})}
\end{equation}
An auxiliary virtual control input $\Lambda$ is formulated to regulate $\zeta_2$ by employing BSC approach. The second tracking error $e_2$ is then defined as:
\begin{equation} \label{eqn43}
    e_2 = \zeta_2 - \Lambda
\end{equation}
Substituting Equation~\eqref{eqn43} into Equation~\eqref{eqn42}, we have 
\begin{equation} \label{eqn44}
    \dot{e}_1 = e_2 + \Lambda + ED_1 - \dot{\zeta}_{1(\text{ref})}
\end{equation}
The Lyapunov control function for $e_1$ is:
\begin{equation} \label{eqn45}
    \Theta_1 = \frac{1}{2} e_1^2
\end{equation}
Its time derivative 
\begin{equation} \label{eqn46}
    \dot{\Theta}_1 = e_1 (e_2 + \Lambda - \dot{\zeta}_{1(\text{ref})}) + e_1 ED_1
\end{equation}
To reinforce stability, the virtual input $\Lambda$ becomes
\begin{equation} \label{eqn47}
    \Lambda = -\beta_1 e_1 + \dot{\zeta}_{1(\text{ref})}
\end{equation}
where $\beta_1>0$ is a user defined constant. Substituting Equation~\eqref{eqn47} into Equation~\eqref{eqn46} simplifies the derivative of Lyapunov control function as:
\begin{equation} \label{eqn48}
    \dot{\Theta}_1 = -\beta_1 e_1^2 + e_1 (e_2 + ED_1)
\end{equation}
For assuring stability, Equation~\eqref{eqn48} must be negative semi-definite at $e_1 = 0$. However, the global stability still must be analyzed. The derivative of the virtual input $\Lambda$ is
\begin{equation} \label{eqn49}
    \dot{\Lambda} = \tau - \gamma
\end{equation}
with $\tau = -\beta_1 (\zeta_2 - \dot{\zeta}_{1(\text{ref})}) + \ddot{\zeta}_{1(\text{ref})}$ and $\gamma = \beta_1 ED_1$. Differentiating $e_2$ from Equation~\eqref{eqn43} yields:
\begin{equation} \label{eqn50}
    \dot{e}_2 = \dot{\zeta}_2 - \dot{\Lambda}
\end{equation}
Using prior equations, we can rewrite:
\begin{equation} \label{eqn51}
    \dot{e}_2 = a(x) + b(x) \mu + ED_2 - \tau + \gamma
\end{equation}

\begin{itemize}
    \item \textbf{Step 2}
\end{itemize}

To ensure robustness, a global integral terminal sliding surface is proposed as~\cite{napole2021global}:
\begin{equation} \label{eqn52}
    \sigma_s = e_2 + \kappa_a \int e_2 \, dt + \kappa_b \left( \int e_2 \, dt \right)^{\frac{x}{y}}
\end{equation}
where $\kappa_a > 0$, $\kappa_b > 0$, and $1 < \frac{x}{y} < 2$. Differentiating $\sigma_s$ yields:
\begin{equation} \label{eqn53}
    \dot{\sigma_s} = \dot{e}_2 + \kappa_a e_2 + \kappa_b (1+\psi) e_2 \left( \int e_2 \, dt \right)^{\psi}
\end{equation}
where $\psi = \frac{x}{y} - 1$. Substituting Equation \eqref{eqn51} gives
\begin{multline} \label{eqn54}
    \dot{\sigma_s} = a(x) + b(x) \mu + ED_2 - \tau + \gamma + \kappa_a e_2 \\
    + \kappa_b (1+\psi) e_2 \left( \int e_2 \, dt \right)^{\psi}
\end{multline}

\begin{itemize}
    \item \textbf{Step 3}
\end{itemize}

Traditional reaching laws suffer from chattering, slow convergence near the sliding surface, and poor robustness to uncertainties. In this paper, we use an improved fractional power-based reaching law (IFPRL) that integrates the traditional constant rate reaching law with the fractional power-based reaching law, defined as follows:

\begin{equation} \label{eqn55}
    \dot{\sigma_s} = -\beta_2 \, \text{sgn}(\sigma_s) - \beta_3 \, |S|^\chi \, \text{sgn}(\sigma_s)
\end{equation}
in which $0 < \chi < 1$, and $\beta_2, \beta_3 > 0$. 

\begin{itemize}
    \item \textbf{Step 4}
\end{itemize}

Equating Equation~\eqref{eqn54} and Equation~\eqref{eqn55}, the control input $\mu$ becomes
\begin{multline} \label{eqn56}
    \mu = \frac{-1}{b(x)} \bigg[ a(x) - \tau + \kappa_a e_2 + \kappa_b (1+\psi) e_2 \left( \int e_2 \, dt \right)^{\psi} \\
    + (\beta_2 + \beta_{\text{ub}1} + \beta_{\text{ub}2}) \, \text{sgn}(\sigma_s) + \beta_3 \, |\sigma_s|^\chi \, \text{sgn}(\sigma_s) \bigg]
\end{multline}
where $|\gamma| \leq \beta_{\text{ub}1}$ and $|ED_2| \leq \beta_{\text{ub}2}$.

\begin{itemize}
    \item \textbf{Step 5}
\end{itemize}

Substituting Equation~\eqref{eqn56} into Equation~\eqref{eqn54}, the sliding surface dynamic becomes
\begin{align} \label{eqn57}
    \dot{\sigma_s} &= -\beta_2 \, \text{sgn}(\sigma_s) - \beta_3 |\sigma_s|^\chi \, \text{sgn}(\sigma_s) 
    - \frac{1}{\sigma_s} e_1 e_2 \nonumber \\
    &\quad - \beta_{\text{ub}1} \, \text{sgn}(\sigma_s) - \beta_{\text{ub}2} \, \text{sgn}(\sigma_s)
\end{align}
Now, we define a composite Lyapunov control function:

\begin{equation} \label{eqn58}
    \Theta_2 = \Theta_1 + \frac{1}{2} \sigma_s^2
\end{equation}
Then, the time derivative of $\Theta_2$ becomes

\begin{multline} \label{eqn59}
    \dot{\Theta}_2 = -\beta_1 e_1^2 - \beta_2 |\sigma_s| - \beta_3 |\sigma_s|^{\chi+1} 
    + |S| (|\gamma| - \\ \beta_{\text{ub}1}) +|\sigma_s| (|ED_2| - \beta_{\text{ub}2})
\end{multline}
By selecting $\beta_{\text{ub}1}$ and $\beta_{\text{ub}2}$ to upper-bound $\gamma$ and $ED_2$, the right-hand side becomes negative semi-definite, ensuring the global stability of the closed-loop system.
%%%%%%%%%

\begin{itemize}
    \item \textbf{Step 6}
\end{itemize}

To confirm finite-time convergence under IFPRL, $\dot{\sigma_s}$ is analyzed in two cases:

\textbf{Case 1:} When $\sigma_s > 0$, $\text{sgn}(\sigma_s) = 1$:
\begin{eqnarray}
\dot{\sigma_s} = -\beta_2 \sigma_s - \beta_3 \sigma_s^\chi = -\sigma_s^\chi (\beta_2 \sigma_s^{1-\chi} + \beta_3)
\end{eqnarray}
The convergence time is
\begin{eqnarray}
t_{\sigma_s>0} = \int_{\sigma_s0}^{0} \frac{d\sigma_s}{-\sigma_s^\chi (\beta_2 \sigma_s^{1-\chi} + \beta_3)} = \frac{\sigma_s0^{1-\chi}}{\beta_3(1-\chi)}
\end{eqnarray}
where $\chi < 1$ ensures finite-time convergence.

\textbf{Case 2:} When $\sigma_s < 0$, $\text{sgn}(\sigma_s) = -1$:
\begin{eqnarray}
\dot{\sigma_s} = \beta_2 \sigma_s - \beta_3 |\sigma_s|^\chi = -|\sigma_s|^\chi (\beta_3 - \beta_2 |\sigma_s|^{1-\chi})
\end{eqnarray}
The convergence time is
\begin{eqnarray}
t_{\sigma_s<0} = \int_{\sigma_s0}^{0} \frac{d\sigma_s}{-|\sigma_s|^\chi (-\beta_2 |\sigma_s|^{1-\chi} + \beta_3)} = \frac{|\sigma_s0|^{1-\chi}}{\beta_3(1-\chi)}
\end{eqnarray}

Thus, for both $\sigma_s>0$ and $\sigma_s<0$, the IFPRL guarantees finite-time convergence of $\sigma_s$ to zero while reducing chattering through the smooth term $\beta_2|\sigma_s|$.

%%%%%%%%%%%%%%%%%%%%%%%%%%%%%
        \begin{figure}[!b]
        \vspace{-10 pt}
        \centering \includegraphics[width= 0.47\textwidth, height = 2.25 in]{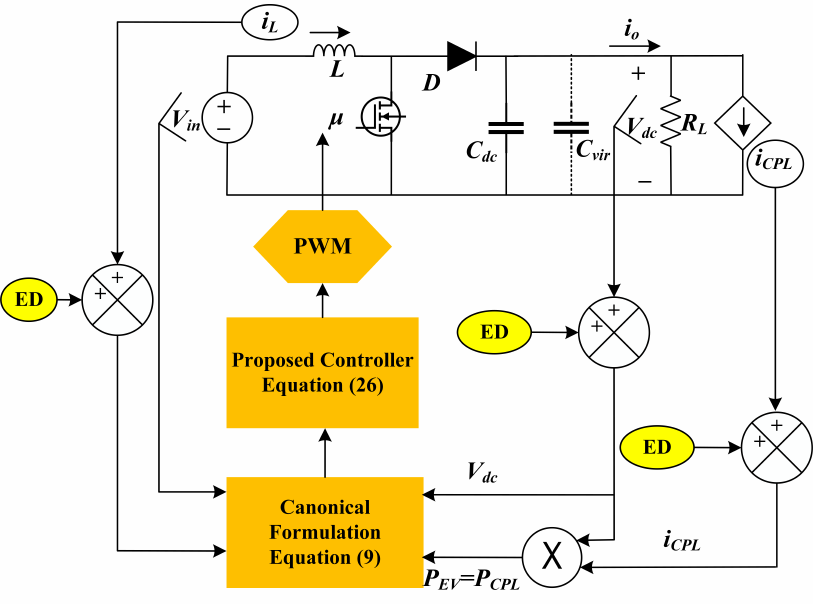}
        \caption{Block diagram of the proposed control strategy for the EV-integrated DCMG.}
        \label{Fig3}
        \end{figure}
%%%%%%%%%%%%%%%%%%%%%%%%%%%%%%%%%%%%

\section{Simulation Results and Discussion}

This section evaluates the performance of the EV-integrated DCMG system through extensive MATLAB/Simulink simulations under realistic scenarios, including time-varying renewable sources, DC-bus voltage reference changes, and control delays. The goal is to validate the effectiveness and robustness of the proposed GITSMBC in regulating DC-bus voltage, preserving dynamic stability, and mitigating nonlinear effects from CPLs. Figure~\ref{Fig3} illustrates the structural realization of the proposed controller within the EV-integrated DCMG system. The setup includes the input source, DBC, feedback linearization-based coordinate transformation, and the control module. Feedback linearization simplifies nonlinear dynamics, enabling faster convergence and higher accuracy. The controller generates the control signal, which is processed by a pulse width modulation (PWM) unit to drive the DBC switches, ensuring stable DC-bus voltage and system reliability. As mention ealier. the EV subsystem is modeled as a CPL which provides NII characteristics that challenge stability. A switching frequency of 10 kHz and a 100 kHz sampling frequency are used to capture transient dynamics with high precision. The control parameters of the proposed GITSMBC scheme are: $\beta_1 = 2000$, $\beta_2 = 50$, $\beta_3 = 5$, $\beta_{ub1} = 1$, $\beta_{ub2} = 1$, $\kappa_a = 0.2$, $\kappa_b = 0.1$, $x = 3$, $y = 5$, $\chi = 0.6$. Table~\ref{tab:parameters} lists the parameters of the EV-integrated DCMG system used in the simulation.

%%%%%%%%%%%%%%%%%%%%%%%%%%%%%%%%%
\begin{table}[t!]
\centering
\caption{Simulation Parameters of the Test System}
\renewcommand{\arraystretch}{1.2}
\begin{tabular}{lcc}
\hline
\textbf{Parameter} & \textbf{Symbol} & \textbf{Value / Unit} \\
\hline
Supply voltage        & $V_{in}$   & 40V/~50V \\
Rated EV power    & $P_{EV}$   & 2.8 kW \\
Converter resistance    & $R$        & 0.07 $\Omega$ \\
Converter inductance    & $L$   & 10.5 $mH$ \\
DC-bus capacitance & $C_{dc}$  & 553.94 $\mu F$ \\
%Load resistance   & $R_L$      & 5.1579 \\
\hline
\end{tabular}
\label{tab:parameters}
\vspace{-15pt}
\end{table}

%%%%%%%%%%%%%%%%%%%%%%%%%%%%%
        \begin{figure}[!b]
        \vspace{-15pt}
        \centering \includegraphics[width= 0.47\textwidth]{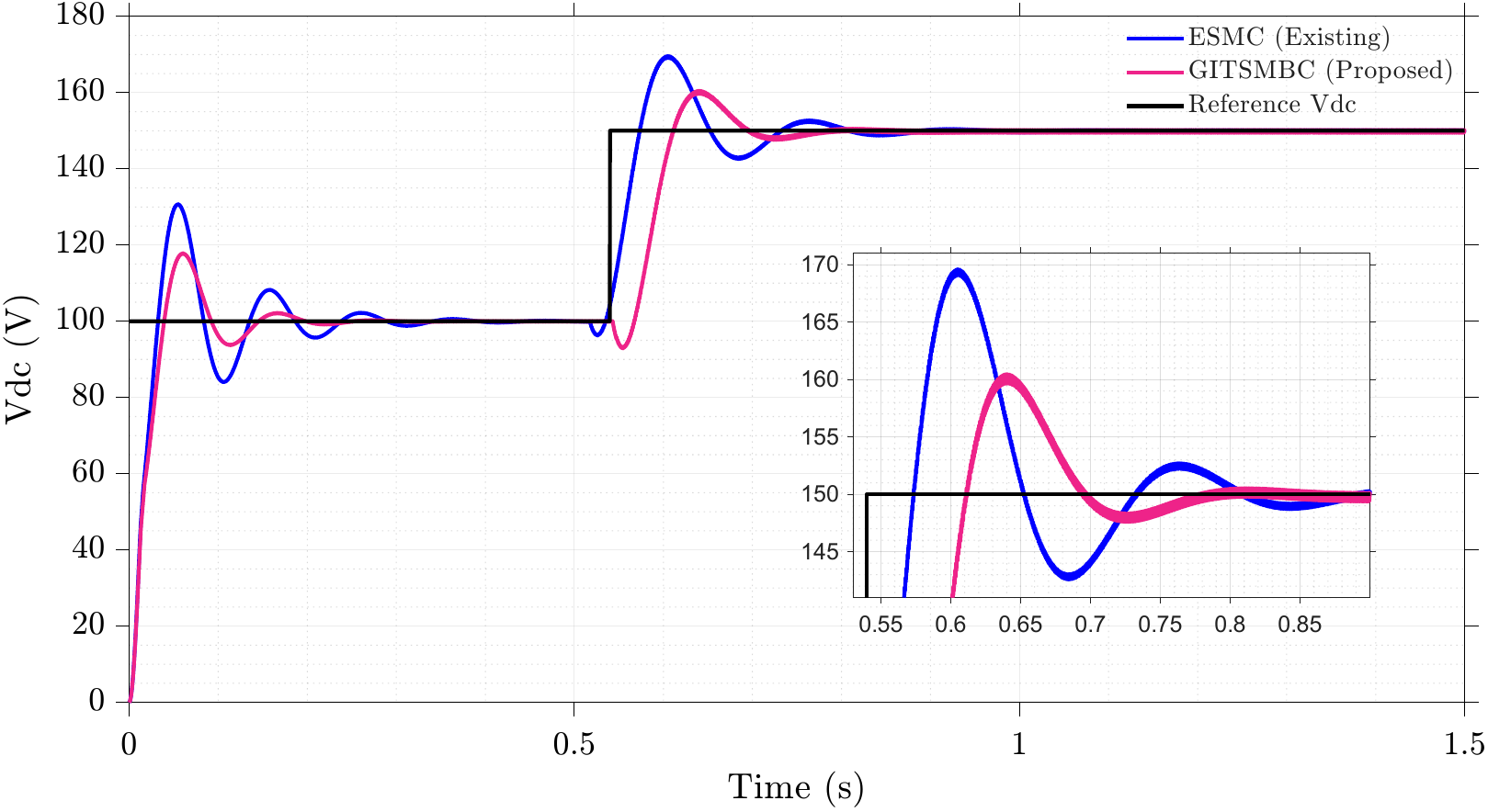}
        \caption{Dynamic response of the DC-bus voltage in scenario 1. At $t = 0$, the proposed GITSMBC reduces overshoot from $30\%$ to $14\%$, undershoot from $15\%$ to $4\%$, and settling time from $38$ ms to $20$ ms compared with ESMC. At $t = 0.54$ s, GITSMBC further improves performance with overshoot of $7\%$ (vs.\ $13.3\%$), undershoot of $0.67\%$ (vs.\ $2.67\%$), and settling time of $28$ ms (vs.\ $43$ ms). }
        \label{Fig_S1}
        \end{figure}
 %%%%%%%%%%%%%%%%%%%%%%%%%%%%%%%%%%%% 

To systematically assess the GITSMBC scheme, two benchmark scenarios are considered: evaluation under varying DC-bus voltage reference (Scenario 1) and evaluation under fluctuating supply voltage with time-varying control delay (Scenario 2). The proposed GITSMBC approach is compared with an existing sliding mode controller (ESMC)~\cite{mohammed2022sliding}, where both controllers are employing virtual capacitors to enhance stability.
%%%%%%%%%%%%%%%%%%%%%%%%%%%%%%%%%%%%%%
\subsection{\textbf{Scenario 1: Evaluation under varying DC-bus voltage reference}}

In this scenario, the supply voltage is set to 50 V, while the EV power is adjusted to 2.8 kW. The reference DC-bus voltage is set to 100 V for the interval $t=0$ to $0.54~s$, and then stepped up to 150 V until $t=1.5~s$.
The dynamic response of the DC-bus under the proposed GITSMBC is illustrated in Figure~\ref{Fig_S1}. As is evident, the controller ensures significantly faster settling time (ST) with negligible overshoot (OS) and undershoot (US), enabling precise and robust tracking of the reference DC-bus voltage. In Figure~\ref{Fig_S1}, the DC-bus voltage response is shown under a time-varying reference. It is observed that the existing controller struggles to accurately follow the reference and exhibits pronounced oscillations, whereas the GITSMBC maintains tight regulation with minimal dynamic deviation. 

From Table~\ref{tab:scenario1}, it is seen that the proposed GITSMBC demonstrates superior transient performance compared to ESMC. At $t = 0$~s, GITSMBC obtains 53.3\% reduction in OS (from 30\% to 14\%), 73.3\% reduction in US (15\% to 4\%), and 47.4\% reduction in ST (38~ms to 20~ms), providing faster DC-bus voltage stabilization. This performance is maintained at $t = 0.54$~s, with further reductions in OS (47.5\%) and US (74.9\%), in conjunction with ~34.9\% faster ST. Although ESMC exhibits a faster initial rise at $0.54$~s, the GITSMBC provides a better overall transient response with small deviation and faster settling.

The overall results in Figure~\ref{Fig_S1} and Table~\ref{tab:scenario1} demonstrate the enhanced regulation, faster response, and superior disturbance rejection capability of the proposed GITSMBC in comparison with the existing approach.

\begin{table}[t!]
\centering
\caption{Transient performance comparison in scenario~1}
\resizebox{\columnwidth}{!}{%
\begin{tabular}{c c c c c c c}
\hline
Time (s) & \multicolumn{3}{c}{ESMC} & \multicolumn{3}{c}{GITSMBC} \\ \cline{2-7}
         & OS (\%) & US (\%) & ST (ms) & OS (\%) & US (\%) & ST (ms) \\ \hline
0        & 30    & 15   & 38  & 14   & 4    & 20 \\ 
0.54      & 13.33 & 2.67 & 43  & 7    & 0.67 & 28 \\ \hline
\end{tabular}%
}
\label{tab:scenario1}
\vspace{-15pt}
\end{table}
%%%%%%%%%%%%%%%%%%%%%%%%%%%%%%%%%%%%%%%
%%%%%%%%%%%%%%%%%%%%%%%%%%%%%
        \begin{figure}[!b]
        \vspace{-15pt}
        \centering \includegraphics[width= 0.47\textwidth]{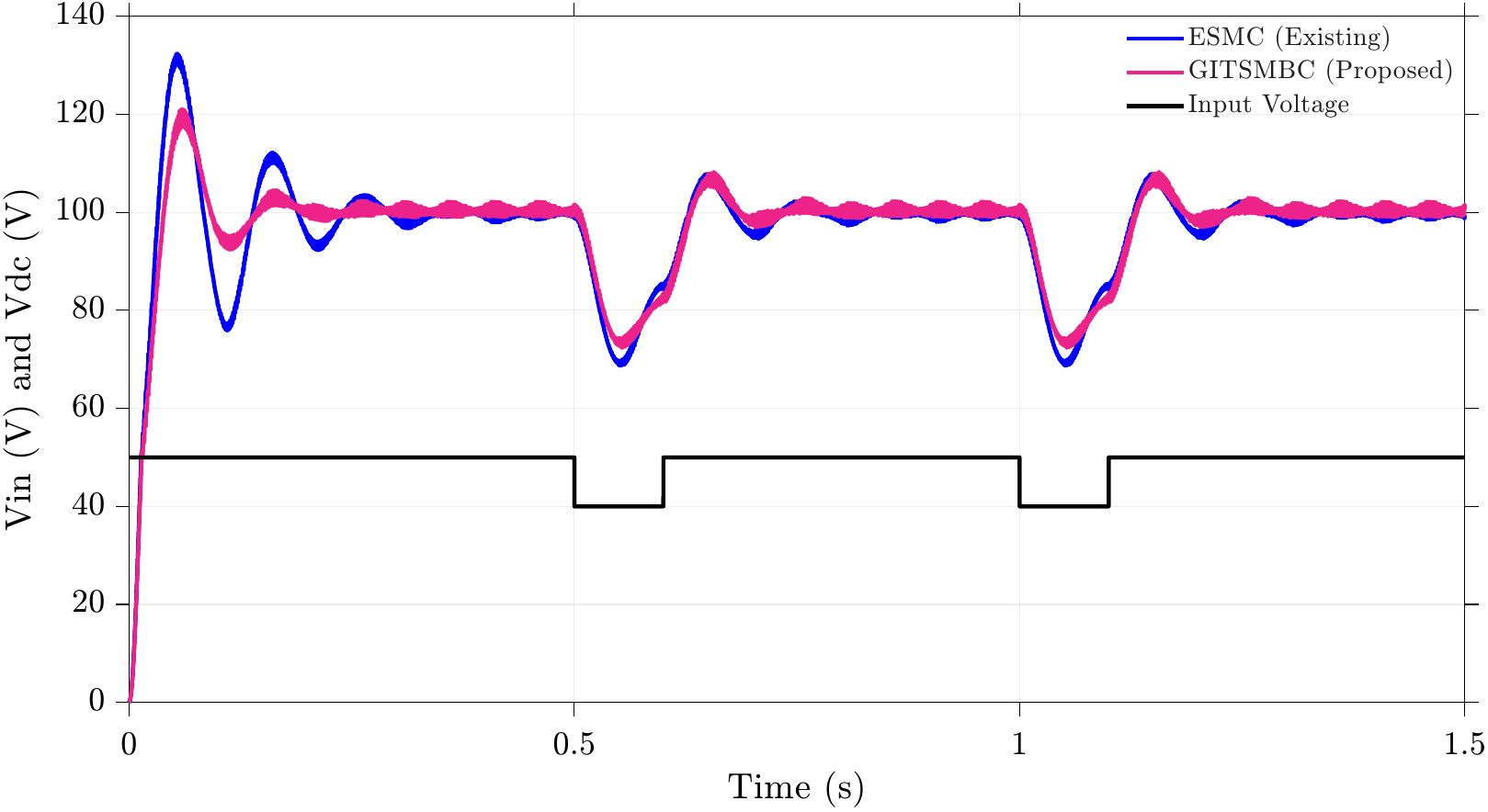}
        \caption{Dynamic response of the DC-bus voltage in scenario 2. At $t = 0$, GITSMBC reduces overshoot from $30.5\%$ to $20\%$, undershoot from $17\%$ to $8\%$, and settling time from $25$ ms to $22$ ms compared with ESMC. At later ($t = 0.5$ s and $t = 1$ s), it maintains lower overshoot and undershoot with comparable settling times.}
        \label{Fig_S2}
        \end{figure}
 %%%%%%%%%%%%%%%%%%%%%%%%%%%%%%%%%%%%

\subsection{\textbf{Scenario 2: Evaluation under fluctuating supply voltage and time-varying control delay}}
To confirm the controller in a realistic cyber-physical system, a time-varying control delay, $t_d = 0.001 + 0.0005 \sin(20\pi t) \text{ s}$ is incorporated. This delay emulates startup latency and network-induced jitter, capturing the joined effects of sensor sampling, computation, and actuator response times. The robustness of the proposed GITSMBC strategy is also verified against the unpredictable nature of renewable sources like solar PV, represented by a variable supply voltage. This combined approach rigorously evaluates the system's performance under practical non-idealities.

The variation in input voltage, illustrated in Figure~\ref{Fig_S2}, is defined as follows: 
\( V_{in} = 50 \, \text{V} \) is maintained for the first \( 0.5 \, \text{s} \). 
At \( t = 0.5 \, \text{s} \), a voltage sag to \( 40 \, \text{V} \) occurs, lasting for \( 0.1 \, \text{s} \), 
before recovering to \( 50 \, \text{V} \). An identical disturbance is applied at \( t = 1.0 \, \text{s} \). Despite substantial supply voltage perturbations, the GITSMBC scheme maintains effective regulation of the DC-bus voltage as depicted in Figure~\ref{Fig_S2}.

Table~\ref{tab:transient_performance} reveals that the proposed GITSMBC scheme enhances transient performance over ESMC across all transient times. At $t = 0$~s, the GITSMBC achieves reductions of 34.4\% in OS, 52.9\% in US, and 12\% in ST. These enhancements are maintained at $t = 0.5$~s (22.2\% OS, 13.3\% US, 11.8\% ST) and $t = 1$~s (16.7\% OS, 10\% US) with ST convergence seen at the latter instant. The outcomes demonstrate that the proposed GITSMBC provides lower deviations and a greater overall transient initially. It is also important to note that the performance metrics, such as US, OS, and ST, are more than those in scenario~1 because of the incorporation of time-varying control delay. 
\vspace{-0.28 em}
\begin{table}[!t]
\centering
\caption{Transient performance comparison in scenario~2}
\resizebox{\columnwidth}{!}{%
\begin{tabular}{ccccccc} 
\hline 
Time (s) & \multicolumn{3}{c}{ESMC} & \multicolumn{3}{c}{GITSMBC} \\ 
\cline{2-7} 
         & OS (\%) & US (\%) & ST (ms) & OS (\%) & US (\%) & ST (ms) \\ 
\hline 
0   & 30.5 & 17   & 25  & 20   & 8   & 22  \\
0.5 & 9    & 30   & 34  & 7    & 26  & 30  \\ 
1   & 9    & 30   & 35  & 7.5  & 27  & 35  \\ 
\hline 
\end{tabular}%
}
\label{tab:transient_performance}
\vspace{-15pt}
\end{table}
%%%%%%%%%%%%%%%%%%%%%%%%%%%%%%%%%%%%

\section{Conclusion and Future Perspectives}
\vspace{-5pt}

This paper presents a robust composite controller to mitigate DC-bus voltage instability in EV-integrated DCMG system, in which the EV acts as a constant power load. The composite control approach combines global integral terminal sliding mode control with backstepping control, embedding a virtual capacitor to emulate inertia and reinforce dynamic performance, and incorporating an improved fractional power reaching law to suppress chattering and enhance transient response. The efficacy of the proposed controller is rigorously evaluated under three challenging conditions: variable reference voltage, variable supply voltage, and a time-varying control delay. A summary of the findings is presented below:

\begin{itemize}
    \item \textbf{Scenario~1 - Voltage reference tracking:} Under step changes in DC-bus voltage, GITSMBC reduces overshoot by up to 53.3\%, undershoot by 74.9\%, and settling time by 47.4\% compared to ESMC.
    
    \item \textbf{Scenario~2 - Robustness to perturbations and delays:} Against supply voltage perturbations and time-varying control delays, GITSMBC decreases overshoot by up to 34.4\%, undershoot by 52.9\%, and settling time by 12\%, demonstrating superior stability under dynamic and cyber-physical conditions.
\end{itemize}

\vspace{-5pt}

\subsection{Future Perspectives} 

\vspace{-4pt} 

In a follow-up work, the controller will be extended with adaptive and self-tuning features to cope with parameter drift, changing loads, and environmental variations in microgrids. Another important direction is enhancing resilience against cyber-physical threats, such as denial-of-service (DoS) and false data injection attacks (FDIAs). Finally, AI-based approaches like reinforcement learning and deep learning will be explored to allow real-time tuning of control parameters, making the system more adaptive, secure, and efficient.
%\vspace{-5pt}
%\section{Artifacts Availability}
%The code, model, and data supporting the results presented in this paper will be made publicly available at \url{https://github.com/AARC-lab/ACC2026} upon acceptance.
%\vspace{-5pt}
\bibliographystyle{IEEEtran}
\bibliography{cas-refs}

\end{document}